\documentclass[twocolumn,showpacs,amsmath,amssymb,aps,
prd,nofootinbib]{revtex4}

\usepackage{graphicx}
\usepackage{amssymb}
\usepackage{amsmath}

\begin{document}

\title{Gravitational wave generation in power-law inflationary models}

\author{Paulo M. S\'a}

%\email{pmsa@ualg.pt}

\affiliation{Centro Multidisciplinar de Astrof\'{\i}sica -- CENTRA
and Departamento de F\'{\i}sica, FCT, Universidade do Algarve,
Campus de Gambelas, 8005-139 Faro, Portugal}

\author{Alfredo B. Henriques}

%\email{alfredo@fisica.ist.utl.pt}

\affiliation{Centro Multidisciplinar de Astrof\'{\i}sica -- CENTRA
and Departamento de F\'{\i}sica, Instituto Superior T\'ecnico,
UTL, Av.\ Rovisco Pais, 1049-001 Lisboa, Portugal}

\date{June 5, 2008}

\begin{abstract}
We investigate the generation of gravitational waves in power-law
inflationary models. The energy spectrum of the gravitational
waves is calculated using the method of continuous Bogoliubov
coefficients. We show that, by looking at the interval of
frequencies between $10^{-5}$ and $10^{5}\mbox{ Hz}$ and also at
the GHz range, important information can be obtained, both about
the inflationary period itself and about the thermalization regime
between the end of inflation and the beginning of the
radiation-dominated era. We thus deem the development of
gravitational wave detectors, covering the MHz/GHz range of
frequencies, to be an important task for the future.
\end{abstract}

 \maketitle

\section{Introduction}

Gravitational waves have the potential of providing us with a
unique telescope and a unique source of information about the very
early universe.

Since the first calculation of the full spectrum of stochastic
gravitational waves has been performed by Allen \cite{allen},
continuing and completing work by other authors, including
Grishchuk \cite{grishchuk} and Starobinsky \cite{starobinsky},
Rubakov, Sazhin and Veryaskin \cite{rubakov-etal}, as well as by
Fabbri and Pollock \cite{fabbri-pollock}, Abbott and Wise
\cite{abbott-wise}, and Abbott and Harari \cite{abbott-harari}, a
lot of research has been done in this direction, particulary into
exploring the potentialities of different inflationary models.
Although the values obtained for the relative energy spectrum of
the cosmological gravitational waves seem, at present, to be
beyond the capabilities of the first generations of
gravitational-wave detectors, this is a situation that is bound to
be changed, given the effort that is being put into the planning
of future detectors with considerably improved sensitivities like,
for instances, NASA's Big Bang Observatory (BBO) \cite{bbo} and
the japanese Deci-hertz Interferometer Gravitational Wave
Observatory (DECIGO) \cite{seto-etal}.

An interesting question to ask is what can we learn about the
physics of the very early universe, from the reading of the
spectrum, assuming an ideal detector. In this article, we address
this question with the help of a popular class of models, namely,
power-law inflation. We shall show that, by looking at the
interval of frequencies between $10^{-5}$ and $10^5 \mbox{ Hz}$,
and also at the GHz range, important information can be obtained,
about both the inflationary period and the thermalization regime.
Concerning the first range of frequencies, between $10^{-5}$ and
$10^5\mbox{ Hz}$, important work, along the lines of our paper,
had already been done \cite{sahni}, albeit by a different method,
with similar conclusions to our own. Most of the developments that
have taken place in gravitational wave detectors cover the first
range of frequencies, but work is also going on to study the
possibility of constructing detectors covering the MHz and GHz
range of frequencies \cite{bernard-etal}. We deem these last
developments to be very important, given the amount of information
that can thereby be obtained, as we show with the help of a simple
model for the transition between the inflationary and
radiation-dominated eras.

In this paper, we consider those gravitational waves generated by
the usual mechanism of quantum fluctuations of the vacuum. In the
MHz/GHz region another independent mechanism is possible, when the
end of inflation is followed by a period of parametric resonance
\cite{traschen-brandenberger}, in which case the gravitational
waves are directly sourced by the large inhomogeneities in the
matter distribution occurring during this period. In this paper we
shall not address this important case, which has been investigated
elsewhere \cite{khlebnikov-tkachev,sa-henriques}, using both
mechanisms.

Throughout, gravitational wave production will be calculated using
the method of continuously evolving Bogoliubov coefficients. This
method, applied to the production of particles in an expanding
universe, was first investigated by Parker \cite{parker}. In the
case of gravitational waves, the differential equations
determining the continuous Bogoliubov coefficients were derived
afterwards in a different, geometrical, way
\cite{moorhouse-henriques-mendes}. The method of continuous
Bogoliubov coefficients has advantages over the frequently used
sudden transition approximation. Associated with the sudden
transition there is always an overproduction of gravitons of large
frequencies, requiring an explicit cut-off for frequencies above
the rate of expansion of the universe \cite{allen}. This cut-off
is obtained in a natural way by the use of continuous Bogoliubov
coefficients \cite{mendes-henriques-moorhouse}. This is also a
very practical method to calculate the full spectrum, from the
very low frequencies corresponding to the present cosmological
horizon, till those large, GHz, frequencies associated with the
transition between inflation and the radiation-dominated universe.

This paper is organized as follows. In Sec.~\ref{sect-power-law}
we introduce the equations defining the power-law inflationary
model. The parameters of the inflationary potential are
constrained using recent measurements from the cosmic microwave
background and large-scale structure. In
Sec.~\ref{sect-reheating}, reheating is incorporated in our model
of evolution through an elementary decay mechanism of the
inflationary scalar field into a relativistic radiation fluid.
Despite its simplicity, our reheating potential may contain
relevant features of more realistic potentials. In
Sec.~\ref{sect-grav-waves} we present the differential equations
to determine the continuous Bogoliubov coefficients and address
the relevant issue of the initial conditions to be used in the
numerical integration of these equations. In
Sec.~\ref{sect-simulations} we describe our numerical simulations
and present several gravitational-wave spectra, obtained for
different values of the parameters of our model. The paper ends
with the Conclusions.

\section{Power-law inflation}\label{sect-power-law}

Let us assume that the evolution of the universe during the
inflationary period is dominated by an homogeneous scalar field
$\phi$ with a potential
\begin{eqnarray}
V(\phi)=V_0 e^{-\lambda\phi}, \label{vphi}
\end{eqnarray}
where $V_0$ and $\lambda$ are positive constants. For a spatially
flat Friedmann-Robertson-Walker metric,
\begin{eqnarray}
d s^{2}=a^2(\eta ) ( -d \eta^2 + d \textbf{x}^2 ),
\label{flat-FRW}
\end{eqnarray}
the Einstein equations for $\phi(\eta)$ and $a(\eta)$ are
\begin{eqnarray}
&& \left( \frac{a^\prime}{a}
\right)^2=\frac{8\pi}{3m_{\textsc{p}}^2} a^2 \left[
\frac{\phi^{\prime 2}}{2a^2} + V(\phi) \right],
\label{friedmann-conforme}
\\
&& \phi^{\prime\prime}+2\frac{a^\prime}{a}\phi^\prime+a^2\frac{d
V}{d \phi}=0, \label{kleingordon-conforme}
\\
&& \frac{a^{\prime\prime}}{a}=-\frac{4\pi}{3m_{\textsc{p}}^2}  a^2
\left[ \frac{\phi^{\prime 2}}{a^2} -4 V(\phi) \right],
\end{eqnarray}
where a prime denotes a derivative with respect to conformal time
$\eta$ and $m_{\textsc{p}}=1/\sqrt{G}=1.22\times10^{19}\mbox{
GeV}$ is the Planck mass.

The above set of differential equations admits the exact solution
\cite{lucchin-matarrese}
\begin{eqnarray}
&& a(\eta) = a_1 (\eta_1-\eta)^p,
\\
&& \phi(\eta) = \frac{1}{\lambda} \ln \left\{
\frac{8\pi}{m_{\textsc{p}}^2} \frac{a_1^2 V_0}{p(2p-1)}
(\eta_1-\eta)^{2(p+1)} \right\},
\end{eqnarray}
where $p<-1$, $a_1$ and $\eta_1$ are arbitrary constants, and
\begin{eqnarray}
\lambda= 4 \sqrt{ \pi \frac{p+1}{p} } m_{\textsc{p}}^{-1}.
\label{lambda}
\end{eqnarray}

Inflationary cosmology predicts a nearly scale-invariant power
spectrum of density perturbations \cite{guth-et-al}. Within the
slow-roll approximation, the power spectrum is given by
\cite{smith-kam-cooray}
\begin{eqnarray}
P_s(k)=P_s(k_c) \left( \frac{k}{k_c}
\right)^{1-n_s+(\alpha_s/2)\ln(k/k_c)}, \label{Ps-1}
\end{eqnarray}
where the scalar power-spectrum amplitude $P_s(k_c)$ and the
spectral parameters $n_s(k_c)$ and $\alpha_s(k_c)$ are evaluated
at some pivot wave number $k_c$. In this paper, we take for these
quantities the values $P_s(k_c)=(2.45\pm0.23)\times10^{-9}$,
$n_s(k_c)=1.0\pm0.1$ and $|\alpha_s(k_c)|<0.04$ at the pivot wave
number $k_c=0.05\mbox{ Mpc}^{-1}$, corresponding to distance
scales of the cosmic microwave background (CMB) and large-scale
structure (LSS) \cite{smith-kam-cooray}.

The spectral parameters $n_s$ and $\alpha_s$ can be defined in
terms of the usual slow-roll parameters $\epsilon$, $\bar{\eta}$
and $\xi$, namely, $n_s=1-6\epsilon+2\bar{\eta}$
\cite{liddle-lyth} and
$\alpha_s=16\epsilon\bar{\eta}-24\epsilon^2-2\xi$
\cite{kosowsky-turner}. For the exponential potential
(\ref{vphi}), the slow-roll parameters are $\epsilon=(p+1)/p$,
$\bar{\eta}=2(p+1)/p$ and $\xi=4(p+1)^2/p^2$, yielding
$n_s=-(p+2)/p$ and $\alpha_s=0$.

Taking into account the above relation between $n_s$ and $p$ and
the CMB/LSS constraint on $n_s$, we conclude that
$p\geqslant-2/1.9\thickapprox-1.053$. Despite the fact that $p$
takes values within a very narrow window, $-1.053\leqslant p<-1$,
it has quite a dramatic effect on the energy spectrum of the
gravitational waves, as will be shown in
Sec.~\ref{sect-simulations}.

Within the slow-roll approximation, the power spectrum
(\ref{Ps-1}) can be written as
\begin{eqnarray}
P_s(k)=\frac{128\pi}{3 m_{\textsc{p}}^6} \left. \frac{V^3}{(d V/d
\phi)^2} \right|_{\phi=\phi_c}= \frac{8V_0
e^{-\lambda\phi_c}}{3m_{\textsc{p}}^4}\frac{p}{p+1}, \label{Ps-2}
\end{eqnarray}
where $\phi_c\equiv\phi(\eta_c)$ is the value of the scalar field
at the moment when the scale $k_c$ exits the Hubble horizon during
the inflationary period. Within our model, $\phi_c$ can be chosen
freely. Let us express it in terms of $\phi_i\equiv\phi(\eta_i)$,
the value of the scalar field at the time reheating begins. Taking
into account that the number of e-foldings of expansion between
$\phi_c$ and $\phi_i$ is given by
\begin{eqnarray}
N_c =\frac{8\pi}{m_{\textsc{p}}^2}\int\limits_{\phi_i}^{\phi_c}
\frac{V}{d V/d \phi} d \phi,
\end{eqnarray}
we obtain
\begin{eqnarray}
\phi_c=\phi_i-N_c\sqrt{\frac{p+1}{4\pi p}}m_{\textsc{p}}.
\end{eqnarray}
Inserting this expression into Eq.~(\ref{Ps-2}), and using
Eq.~(\ref{Ps-1}), we obtain a constraint on the inflationary scale
$V_0$, namely,
\begin{eqnarray}
\hspace{-5mm} V_0&=&\frac{3(p+1)}{8p} P_s(k_c) \nonumber \\
\hspace{-5mm} && \times \exp \left\{
4\sqrt{\pi\frac{p+1}{p}}\frac{\phi_i}{m_{\textsc{p}}} -2N_c
\frac{p+1}{p}\right\} m_{\textsc{p}}^4, \label{V0}
\end{eqnarray}
where $47\lesssim N_c \lesssim 62$ \cite{smith-kam-cooray} and, as
already mentioned above, $\phi_i$ is a free parameter, marking the
beginning of the reheating process. In the numerical simulations,
described in detail in Sec.~\ref{sect-simulations}, we will use
$\phi_i=-0.3 m_{\textsc{p}}$, $P_s(k_c)=2.45\times10^{-9}$ and
$N_c=55$, implying that $V_0^{1/4}\leqslant1.5\times10^{16}\mbox{
GeV}$ for $-1.053\leqslant p<-1$.

\section{Reheating}\label{sect-reheating}

\begin{figure}[t]
\includegraphics[width=8.6cm]{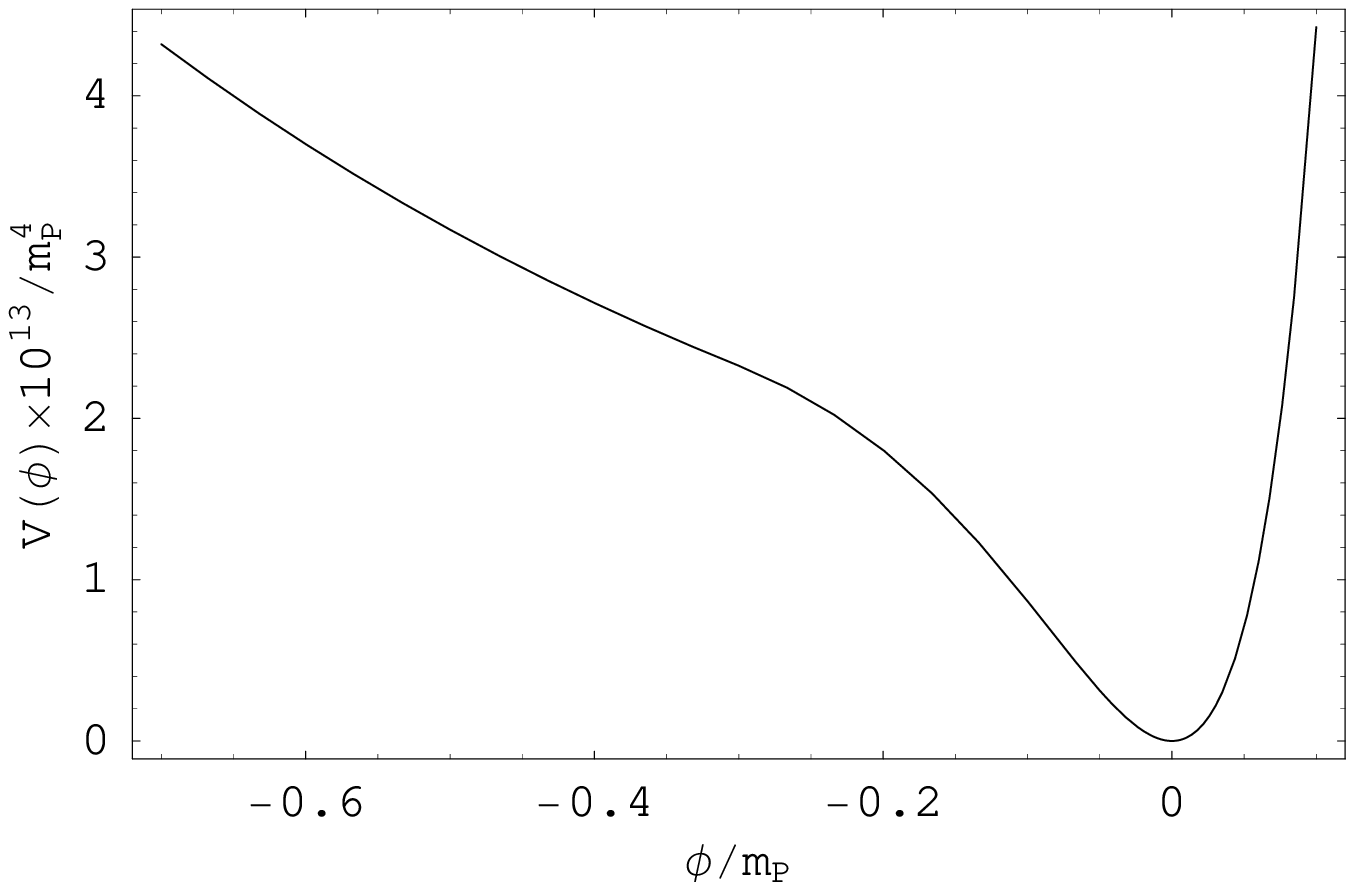}
\caption{Potential $V(\phi)$ for $p=-1.05$, $\phi_i=-0.3
m_{\textsc{p}}$ and $n=2$, implying $\lambda=1.547
m_{\textsc{p}}^{-1}$, $V_0=1.461\times10^{-13}m_{\textsc{p}}^4$,
$U_0=2.785\times10^{-13}m_{\textsc{p}}^4$ and $\nu=8.152
m_{\textsc{p}}^{-1}$.} \label{fig-potential}
\end{figure}

In order to incorporate reheating in our model of evolution, we
assume that the potential of the scalar field is no longer given
by Eq.~(\ref{vphi}), but rather by
\begin{eqnarray}
V(\phi) = \left\{
\begin{tabular}{ll}
 $V_0 e^{-\lambda \phi}$, & \quad $\phi \leqslant \phi_i$, \\
 $U_0 \left( e^{\nu\phi}-1 \right)^n$,  & \quad
 $\phi\geqslant\phi_i$,
\end{tabular}
\right. \label{reheating-potential}
\end{eqnarray}
where $n$ is a even number and $U_0$ and $\nu$ are chosen such
that the potential and its first derivative are continuous at
$\phi=\phi_i$ (see Fig.~\ref{fig-potential}).

This is clearly a toy potential. However, it may contain relevant
features of more realistic potentials, arising from extensions of
the standard model; these are likely to involve potentials defined
by power-series expansions of scalar fields, or by exponentials
\cite{lyth}. An important example of the last type, may be seen in
the application of the Salam-Sezgin six-dimensional supergravity
model \cite{salam-sezgin} to cosmology \cite{anchordoqui-etal}.
Our potential is similar to the one derived within the
Salam-Sezgin model, if in Eq.~(\ref{reheating-potential}) we put
n=2.

Initially, the scalar field $\phi$ has a large negative value. A
period of pure power-law inflation takes place until $\phi$ rolls
down to values of the order of $\phi_i$. After that, the scalar
field rapidly approaches the minimum of the potential, where $V
\approx U_0 \nu^n\phi^n$, beginning to oscillate around it. The
energy of the scalar field is then transferred to a relativistic
radiation fluid and reheating takes place. The decay rate of the
scalar field into the radiation fluid is governed by a
phenomenological parameter, $\Gamma_\phi$, which we choose to be
of the order of $10^{-7}m_{\textsc{p}}-10^{-8}m_{\textsc{p}}$.

During the period of transition between power-law inflation and
the radia\-tion-dominated era, the evolution of the universe is
described by the set of differential equations
\begin{eqnarray}
\hspace{-7mm} && \left( \frac{a^\prime}{a} \right)^2 =
\frac{8\pi}{3m_{\textsc{p}}^2} a^2 \left[
\frac{\phi^{\prime2}}{2a^2} + U_0 \left( e^{\nu\phi}-1 \right)^n +
\rho_{\mbox{\scriptsize rad}}
\right], \label{friedmann1} \\
\hspace{-7mm} &&  \frac{a^{\prime\prime}}{a} =
-\frac{4\pi}{3m_{\textsc{p}}^2} a^2 \left[
\frac{\phi^{\prime2}}{a^2} - 4 U_0 \left( e^{\nu\phi}-1 \right)^n
\right],\label{adotdot1}
\\
\hspace{-7mm} && \phi^{\prime\prime} +2 \frac{a'}{a} \phi' + a^2 n
\nu U_0 \left( e^{\nu\phi}-1 \right)^{n-1} e^{\nu\phi} =  -
\Gamma_{\phi} a \phi^\prime,
\\
\hspace{-7mm} && \rho_{\mbox{\scriptsize rad}}^\prime
+4\frac{a^\prime}{a} \rho_{\mbox{\scriptsize rad}} = \Gamma_\phi
\frac{\phi'^2}{a}, \label{rho1}
\end{eqnarray}
where $\rho_{\mbox{\scriptsize rad}}$ is the energy density of the
relativistic radiation fluid.

The transition period ends when the energy density of the scalar
field, $\rho_\phi=\phi^{\prime2}/(2a^2)+U_0(e^{\nu\phi}-1)^n$,
becomes much smaller than the energy density of radiation,
$\rho_{\mbox{\scriptsize rad}}$. After that, we neglect the scalar
field and the evolution of the universe till the present time is
described by the set of differential equations
\begin{eqnarray}
\frac{a^{\prime\prime}}{a} &=& \frac{4\pi G}{3} a^2 \left[
\rho_{\mbox{\scriptsize mat},0} \left( \frac{a_0}{a} \right)^3
\right.
\nonumber \\
&& \left. +\, (3w+1)\rho_{\mbox{\scriptsize de},0} \left(
\frac{a_0}{a} \right)^{3(1-w)} \right], \label{101}
\\
\left( \frac{a^{\prime}}{a} \right)^2 &=& \frac{8\pi G}{3} a^2
\left[ \rho_{\mbox{\scriptsize rad},0} \left( \frac{a_0}{a}
\right)^4 + \rho_{\mbox{\scriptsize mat},0} \left( \frac{a_0}{a}
\right)^3 \right. \nonumber \\
&& \left. +\, \rho_{\mbox{\scriptsize de},0} \left( \frac{a_0}{a}
\right)^{3(1-w)} \right], \label{102}
\end{eqnarray}
where $w=0.78$ and the density of radiation, matter and dark
energy at the present time $\eta_0$ are, respectively,
$\rho_{\mbox{\scriptsize rad},0}=4.6\times 10^{-31} \mbox{
kg/m}^3$, $\rho_{\mbox{\scriptsize mat},0}=2.6\times10^{-27}
\mbox{ kg/m}^3$ and $\rho_{\mbox{\scriptsize
de},0}=6.9\times10^{-27} \mbox{ kg/m}^3$, yielding
$H(\eta_0)=71\mbox{ km s}^{-1}\mbox{Mpc}^{-1}$.

\section{Gravitational-wave spectrum}\label{sect-grav-waves}

The dimensionless relative logarithmic energy spectrum of the
gravitational waves at the present time $\eta_0$ is defined by
\begin{eqnarray}
\Omega_{\mbox{\scriptsize GW}} (\eta_0) &=&
\frac{1}{\rho_{\mbox{\scriptsize crit}}(\eta_0)}
\frac{d\rho_{\mbox{\scriptsize GW}}}{d \ln\omega}(\eta_0)
\nonumber \\
&=& \frac{8\hbar G}{3\pi c^5 H^2(\eta_0)} \omega^4 (\eta_0)
\beta^2(\eta_0), \label{8}
\end{eqnarray}
where $\rho _{\mbox{\scriptsize crit}}$ is the critical density of
the universe, $\rho _{\mbox{\scriptsize GW}}$ and $\omega$ are the
energy density and angular frequency of the gravitational waves,
respectively, $H$ is the Hubble parameter and $\beta$ is a
Bogoliubov coefficient, such that $|\beta|^2=\langle
N_k(\eta)\rangle$ gives the number of gravitons.

In order to calculate the amount of gravitons produced during the
evolution of the universe we use Parker's method of continuous
Bogoliubov coefficients \cite{parker,moorhouse-henriques-mendes},
according to which the Bogoliubov coefficients $\alpha(\eta)$ and
$\beta(\eta)$, defined as continuous functions of time, are
determined by the set of differential equations
\begin{eqnarray}
&& \alpha^\prime = \frac{i}{2k}\left[ \alpha+\beta e^{2i
k(\eta-\eta_i)} \right] \frac{a^{\prime\prime}}{a},
\label{bogo-alpha}
\\
&& \beta^\prime = -\frac{i}{2k}\left[ \beta+\alpha e^{-2i
k(\eta-\eta_i)} \right] \frac{a^{\prime\prime}}{a},
\label{bogo-beta}
\end{eqnarray}
where $k=2\pi a(\eta)/\lambda(\eta) =a(\eta)\omega(\eta)$ is the
co-moving wave number. This method has been used by us in previous
works to investigate the generation of gravitational waves in
different cosmological models
\cite{sa-henriques,mendes-henriques-moorhouse,henriques}.

Upon the redefinition
\begin{eqnarray}
&& \alpha = \frac12 (X+Y) e^{i k(\eta -\eta_i)},\\
&& \beta = \frac12 (X-Y) e^{-i k(\eta -\eta_i)},
\label{redefinition}
\end{eqnarray}
the system of Eqs.~(\ref{bogo-alpha}) and (\ref{bogo-beta}) become
simply
\begin{eqnarray}
&& X^{\prime\prime} +\left( k^{2}-\frac{a^{\prime \prime }}{a}
\right) X=0, \label{6}
\\
&& Y = \frac{i}{k} X^\prime. \label{6a}
\end{eqnarray}

These differential equations can be solved exactly for pure
power-law inflation. Indeed, for $a(\eta)\propto (\eta_1-\eta)^p$,
Eq.~(\ref{6}) can be written as
\begin{eqnarray}
 \frac{d^2 X}{d y^2}+\left( 1-\frac{4\mu^2-1}{4y^2} \right) X=0, \label{eq-X-y}
\end{eqnarray}
which admits the solution
\begin{eqnarray}
X(y)=y^{1/2} \left[ c_1 J_\mu(y)+c_2 Y_\mu (y)\right],
\end{eqnarray}
where $y=k(\eta_1-\eta)$, $\mu=\pm(1/2-p)$, and $J_\mu$ and
$Y_\mu$ are Bessel functions of the first and the second kind.

The functions $X$ and $Y$ should satisfy the initial conditions
$|X(\eta=-\infty)|=1$ and $|Y(\eta=-\infty)|=1$, corresponding to
$|\alpha(\eta=-\infty)|=1$ and $|\beta(\eta=-\infty)|=0$. This
implies that the integration constants should be $c_1 =
\sqrt{\pi/2} e^{i\delta}$ and $c_2 =\pm i \sqrt{\pi/2}
e^{i\delta}$, where $\delta$ is an arbitrary real constant. The
solution of Eq.~(\ref{6}), which satisfies appropriate initial
conditions, is then
\begin{eqnarray}
X(y)= \sqrt{\frac{\pi y}{2}} H_{1/2-p}^{(1,2)} (y) e^{i\delta},
\end{eqnarray}
where $H_{1/2-p}^{(1,2)}=J_{1/2-p}\pm i Y_{1/2-p}$ denotes the
Hankel functions.

From Eq.~(\ref{6a}) is now straightforward to obtain
\begin{eqnarray}
Y(y)= -i\sqrt{\frac{\pi y}{2}} \left[ H_{-1/2-p}^{(1,2)} (y)+
\frac{p}{y} H_{1/2-p}^{(1,2)} (y) \right]e^{i\delta},
\end{eqnarray}
which satisfies the initial condition $|Y(\eta=-\infty)|=1$.

At the end of the pure power-law inflationary period, where our
numerical integration starts ($\eta=\eta_i$), the functions $X$
and $Y$ are given by
\begin{eqnarray}
\hspace{-9mm} && X(\eta_i)= \sqrt{\frac{\pi y_i}{2} }
H_{1/2-p}^{(1,2)} (y_i), \label{Xi}
\\
\hspace{-9mm} && Y(\eta_i)= -i\sqrt{\frac{\pi y_i}{2}} \left[
H_{-1/2-p}^{(1,2)} (y_i)+ \frac{p}{y_i} H_{1/2-p}^{(1,2)} (y_i)
\right], \label{Yi}
\end{eqnarray}
where $y_i=k(\eta_1-\eta_i)=-k p\, a(\eta_i)/a^\prime(\eta_i)$ and
we have chosen the arbitrary constant $\delta$ to be zero.
Equations (\ref{Xi}) and (\ref{Yi}) will be used as initial
conditions for $X$ and $Y$ in our numerical simulations.

\section{Numerical simulations}\label{sect-simulations}

The numerical simulations begin at the end of the pure power-law
inflationary period, $\eta_i$, and continue up to the present
epoch, $\eta_0$. We use a Runge-Kutte method to solve the system
of differential equations (\ref{adotdot1})--(\ref{101}). Equations
(\ref{friedmann1}) and (\ref{102}) are used as constraint
equations to check the accuracy of the numerical solution.

As initial conditions we choose $a(\eta_i)=1$ and
$\phi(\eta_i)\equiv\phi_i= -0.3 \, m_{\textsc{p}}$. Taking into
account that, for $\phi\leqslant\phi_i$, $\phi^{\prime
2}/(2a^2)=V(\phi)(p+1)/(2p-1)$, we obtain
\begin{eqnarray}
\phi^\prime(\eta_i)=\sqrt{\frac{2(p+1)}{2p-1} V_0
e^{-\lambda\phi_i}}, \label{phiprimeinitial}
\end{eqnarray}
where $\lambda$ and $V_0$ are given by Eqs.~(\ref{lambda}) and
(\ref{V0}), respectively. Since any pre-existing radiation fluid
is diluted during inflation, we choose $\rho_{\mbox{\scriptsize
rad}}(\eta_i)=0$. Finally, $a^\prime(\eta_i)$ is determined from
the constraint (\ref{friedmann1}) to be
\begin{eqnarray}
a^\prime(\eta_i)=\sqrt{3\pi \frac{p+1}{2p-1} P_s(k_c)}
e^{-N_c(p+1)/p} \; t_{\mbox{\scriptsize P}}^{-1}, \label{aprime-i}
\end{eqnarray}
where $t_{\mbox{\scriptsize P}}$ is the Planck time and we have
taken into account that $U_0(e^{\nu\phi_i}-1)^n=V_0
e^{-\lambda\phi_i}$.

Having determined the time evolution of $a^{\prime\prime}/a$ from
Eqs.~(\ref{adotdot1})--(\ref{101}), we then solve Eqs.~(\ref{6})
and (\ref{6a}) (again with a Runge-Kutte method) for different
values of $\omega(\eta_0)=k/a(\eta_0)$ and with initial conditions
(\ref{Xi}) and (\ref{Yi}). Finally, we compute $\beta^2$ with
Eq.~(\ref{redefinition}) and obtain $\Omega_{\mbox{\scriptsize
GW}}(\eta_0)$ from Eq.~(\ref{8}).

The gravitational-wave energy spectrum can be divided in three
regions. In the first region, $\omega_0\equiv\omega(\eta_0)$
ranges from $\omega_{\mbox{\scriptsize min}}$ to
$\omega_{\mbox{\scriptsize rad} \rightarrow \mbox{\scriptsize
mat}}$, where $\omega_{\mbox{\scriptsize min}}=2\pi
c/d_{\mbox{\scriptsize Hubble}}(\eta_0)\approx 2\pi
H(\eta_0)=1.4\times10^{-17} \mbox{ rad/s}$ is the angular
frequency of a gravitational wave with a wavelength equal, today,
to the Hubble distance, and $\omega_{\mbox{\scriptsize rad}
\rightarrow \mbox{\scriptsize mat}}$ is today's value of the
angular frequency of a gravitational wave which had a wavelength
equal to the Hubble distance at the time when the energy density
of radiation became equal to the energy density of matter
(corresponding to the transition from a radiation-dominated to a
matter-dominated universe). Because of red-shifting,
$\omega_{\mbox{\scriptsize rad} \rightarrow \mbox{\scriptsize
mat}}$ is of the order of $10^{-15} \mbox{ rad/s}$. In the second
(intermediate) region, $\omega_0$ ranges from
$\omega_{\mbox{\scriptsize rad} \rightarrow \mbox{\scriptsize
mat}}$ to $\omega_{\phi\rightarrow\mbox{\scriptsize rad}}$, where
$\omega_{\phi\rightarrow\mbox{\scriptsize rad}}$ is today's value
of the angular frequency of a gravitational wave which had a
wavelength equal to the Hubble distance at the time when the
energy density of the inflaton became equal to the energy density
of radiation (marking the end of the transition period between the
inflationary and radiation-dominated eras); this frequency is of
the order of $10^7\mbox{ rad/s}$. Finally, in the third region,
$\omega_0$ ranges from $\omega_{\phi\rightarrow\mbox{\scriptsize
rad}}$ to $\omega_{\mbox{\scriptsize max}}$, where the maximum
angular frequency, which is of the order of $10^{10}\mbox{
rad/s}$, is today's value of the angular frequency of a
gravitational wave which had a wavelength equal to the Hubble
distance at the end of the inflationary period.

Let us now show, using the method of continuous Bogoliubov
coefficients, that in the second (intermediate) region of the
spectrum, the relative logarithmic energy density of the
gravitational waves, contrarily to the situation in exponential
inflation, depends on the frequency of the waves, namely,
$\Omega_{\mbox{\scriptsize GW}} \propto \omega_0^{2(p+1)}$.

Taking into account that, according to our numerical simulations,
the scale factor at the present time is of the order of $10^{30}$
[for $a(\eta_i)=1$], and using Eq.~(\ref{aprime-i}) to eva\-luate
$a^\prime(\eta_i)$, we obtain
\begin{eqnarray}
y_i \approx -3.5 \times 10^{-10} p \sqrt{\frac{2p-1}{p+1}}
e^{55(p+1)/p} \; \omega_0\mbox{ s}. \label{yi}
\end{eqnarray}
Let us consider values of $p$ in the interval $-1.053\leqslant
p\leqslant-1-\epsilon$. From Eq.~(\ref{yi}) follows that $y_i<1$
for $\omega_0 \leqslant 2.2 \times 10^7 \mbox{ rad/s}$ and
$\epsilon= 1.9\times10^{-4}$, allowing us to expand $X(\eta_i)$
and $Y(\eta_i)$, given by Eqs.~(\ref{Xi}) and (\ref{Yi}), in
converging power series. Keeping just the leading term,
$X(\eta_i)\propto y_i^p$ and $Y(\eta)\propto y_i^{p-1}$, we obtain
that $\beta^2(\eta_i)\propto y_i^{2(p-1)}\propto
\omega_0^{2(p-1)}$.

\begin{figure}[t]
\includegraphics[width=8.6cm]{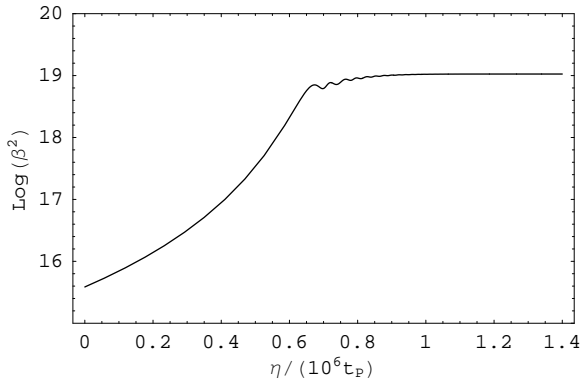}
\caption{Time evolution of $\beta^2$ during the transition from
the inflationary period to the radiation-dominated era, for
$\omega_0=10^4\mbox{ rad/s}$, $p=-1.05$,
$\phi_i=-0.3m_{\textsc{p}}$,
$\Gamma_\phi=5.0\times10^{-7}m_{\textsc{p}}$ and $n=2$.}
\label{fig-beta}
\end{figure}

In order to determine the subsequent time evolution of $\beta^2$
we have to solve Eq.~(\ref{6}). Let us consider separately three
periods: \textit{i}) the transition from inflation to the
radiation-dominated era ($\eta_i \leqslant \eta <
\eta_{\phi\rightarrow\mbox{\scriptsize rad}}$) \textit{ii}) the
radiation-dominated era ($\eta_{\phi\rightarrow\mbox{\scriptsize
rad}} \leqslant \eta < \eta_{\mbox{\scriptsize rad} \rightarrow
\mbox{\scriptsize mat}}$) and \textit{iii}) the matter- and dark
energy-dominated eras ($\eta_{\mbox{\scriptsize rad} \rightarrow
\mbox{\scriptsize mat}} \leqslant \eta \leqslant \eta_0$).

During the transition from inflation to the radiation-dominated
era, $a^{\prime\prime}/a$ is a rather complicated function of time
and, therefore, Eq.~(\ref{6}) must be solved numerically. In
Fig.~\ref{fig-beta} the time evolution of $\beta^2$ is shown for
$\omega_0=10^4\mbox{ rad/s}$, $p=-1.05$,
$\phi_i=-0.3m_{\textsc{p}}$,
$\Gamma_\phi=5.0\times10^{-7}m_{\textsc{p}}$ and $n=2$. For other
values of $\omega_0$, the time evolution of $\beta^2$ is the same,
provided that $\omega_0 \leqslant 2.2 \times 10^7 \mbox{ rad/s}$.
Indeed, for such values of $\omega_0$, $k^2 = a^2_0 \omega^2_0 <
1.4 \times 10^{-12} \, t_{\mbox{\scriptsize P}}^{-2}$ is always
smaller than $a^{\prime\prime}(\eta_i)/a(\eta_i)$. To see this, we
insert $\phi^\prime(\eta_i)$ from Eq.~(\ref{phiprimeinitial}) into
Eq.~(\ref{adotdot1}), use the fact that
$U_0(e^{\nu\phi_i}-1)^n=V_0 e^{-\lambda\phi_i}$, with $\lambda$
and $V_0$ given by Eqs.~(\ref{lambda}) and (\ref{V0}), obtaining,
for $\epsilon= 1.9\times10^{-4}$,
\begin{eqnarray}
\frac{a^{\prime\prime}}{a}(\eta_i) &=& 3\pi
\frac{p^2-1}{p(2p-1)}P_s(k_c) e^{-2N_c(p+1)/p} \;
t_{\mbox{\scriptsize P}}^{-2} \nonumber \\
&>& 2.9\times10^{-12} \, t_{\mbox{\scriptsize P}}^{-2}.
\end{eqnarray}
In the first stages of the transition between the inflationary and
radiation-dominated eras, $a^{\prime\prime}/a$ increases and the
relation $k^2<a^{\prime\prime}/a$ continues to hold. Therefore, we
can neglect $k^2$ in Eq.~(\ref{6}), thus obtaining the same time
evolution of $\beta^2$ for different values of $\omega_0$. To the
end of the transition, $a^{\prime\prime}/a$ approaches zero and
graviton production ceases, with $\beta^2$ becoming constant (see
Fig.~\ref{fig-beta}).

During the radiation-dominated era, $\beta^2$ remains constant.
Indeed, during this era the scale factor is proportional to the
conformal time, implying $a^{\prime\prime}/a=0$. Then
Eqs.~(\ref{6}) and (\ref{6a}) admit the solution
\begin{eqnarray}
\hspace{-9mm} && X(\eta) =
X(\eta_{\phi\rightarrow\mbox{\scriptsize rad}}) \cos q(\eta) -i
Y(\eta_{\phi\rightarrow\mbox{\scriptsize rad}})\sin q(\eta),
\\
\hspace{-9mm} && Y(\eta) =
Y(\eta_{\phi\rightarrow\mbox{\scriptsize rad}}) \cos q(\eta) -i
X(\eta_{\phi\rightarrow\mbox{\scriptsize rad}})\sin q(\eta),
\end{eqnarray}
where $q(\eta) \equiv  k
(\eta-\eta_{\phi\rightarrow\mbox{\scriptsize rad}})$. Now, using
Eq.~(\ref{redefinition}), it is straightforward to show that
$\beta^2(\eta)=\beta^2(\eta_{\phi\rightarrow\mbox{\scriptsize
rad}})$.

\begin{figure}[t]
\includegraphics[width=8.6cm]{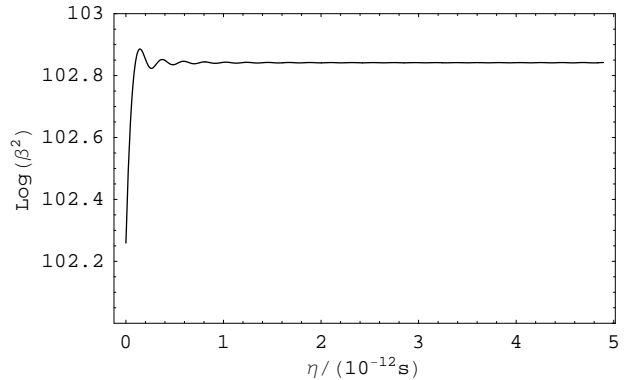}
\caption{Time evolution of $\beta^2$ during the matter- and dark
energy-dominated eras, for $\omega_0=5\times10^{-17}\mbox{
rad/s}$, $p=-1.05$, $\phi_i=-0.3m_{\textsc{p}}$,
$\Gamma_\phi=5.0\times10^{-7}m_{\textsc{p}}$ and $n=2$.}
\label{fig-beta-2}
\end{figure}

During the matter- and dark energy-dominated eras, $\beta^2$
remains constant, if $\omega_0\gtrsim10^{-15}\mbox{ rad/s}$.
Indeed, for such values of $\omega_0$, our numerical simulations
show that $k^2$ is always greater than $a^{\prime\prime}/a$ and,
once again, we can neglect $a^{\prime\prime}/a$ in Eq.~(\ref{6}),
obtaining, similarly to the radiation-dominated era, that
$\beta^2$ remains constant, i.e.,
$\beta^2(\eta)=\beta^2(\eta_{\mbox{\scriptsize
rad}\rightarrow\mbox{\scriptsize mat}})$.

Putting everything together, we arrive at the conclusion that, for
$10^{-15} \mbox{ rad/s} \lesssim \omega_0 \lesssim 10^{7}\mbox{
rad/s}$ (intermediate region of the gravitational-wave spectrum),
gravitons are produced during the transition from the inflationary
period to the radiation-dominated era, but not during the
radiation-, matter-, and dark energy-dominated eras, and that this
production does not depend on the value of $\omega_0$. This,
together with the fact that $\beta^2(\eta_i) \propto
\omega_0^{2(p-1)}$, implies that $\beta^2(\eta_0)\propto
\omega_0^{2(p-1)}$. Now, using Eq.~(\ref{8}), we finally obtain
that $\Omega_{\mbox{\scriptsize GW}}(\eta_0) \propto
\omega_0^{2(p+1)}$, i.e., for $10^{-15}\mbox{ rad/s} \lesssim
\omega_0 \lesssim 10^{7}\mbox{ rad/s}$ the relative logarithmic
energy density of the gravitational waves depends on the frequency
of the waves, in agreement with the results obtained by Sahni
using a different method \cite{sahni}. This contrasts with the
results obtained for exponential inflation, where the spectrum in
the intermediate region is flat.

Let us point out that for $1.4\times10^{-17}\mbox{ rad/s} \lesssim
\omega_0 \lesssim 10^{-15}\mbox{ rad/s}$, the energy spectrum is
steeper than in the intermediate-frequency region. This is due to
the fact that, in the beginning of the matter-dominated era,
$a^{\prime\prime}/a$ is greater than $k^2$ for such range of
frequencies, leading to an extra production of gravitons. After a
while, $a^{\prime\prime}/a$ approaches zero and graviton
production ceases, with $\beta^2$ becoming constant (see
Fig.~\ref{fig-beta-2}).

\begin{figure}[t]
\includegraphics[width=8.6cm]{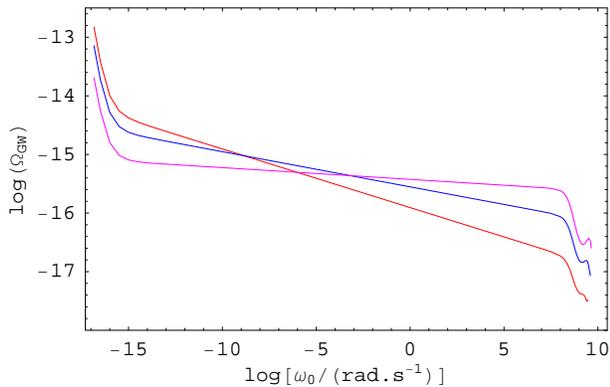}
\caption{Gravitational-wave spectra for different values of $p$:
$p=-1.01$ (violet, upper curve at high frequencies), $p=-1.03$
(blue, middle curve) and $p=-1.05$ (red, lower curve). In all
spectra we have used $\phi_i=-0.3m_{\textsc{p}}$,
$\Gamma_\phi=5.0\times10^{-7}m_{\textsc{p}}$ and $n=2$.}
\label{fig-varios-p}
\end{figure}

We now produce a few gravitational-wave spectra, coming from our
numerical simulations. For each spectrum, we have to specify the
values of $p$, $\phi_i$, $n$ and $\Gamma_\phi$. While $p$ is
constrained by measurements of the cosmic microwave background and
large-scale structure to lie in the interval $-1.053\leqslant
p<-1$, the latter three parameters can be chosen freely. Our
choice of $p$ fixes $V_0$ and $\lambda$ and, thus, the level of
the spectrum.

\begin{figure}[t]
\includegraphics[width=8.6cm]{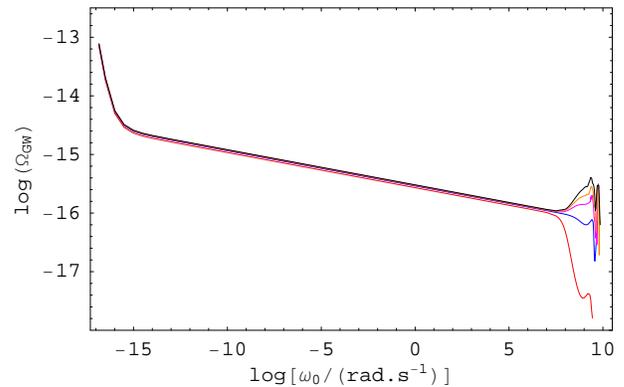}
\caption{Gravitational-wave spectra for different values of $n$:
$n=2$ (red, lower curve), $n=4$ (blue, second curve from below),
$n=6$ (violet, third curve from below), $n=8$ (orange, second
curve from above) and $n=12$ (black, upper curve). In all spectra
we have used $p=-1.03$, $\phi_i=-0.3m_{\textsc{p}}$ and
$\Gamma_\phi=5.0\times10^{-8}m_{\textsc{p}}$.}
\label{fig-varios-n}
\end{figure}

\begin{figure}[t]
\includegraphics[width=8.6cm]{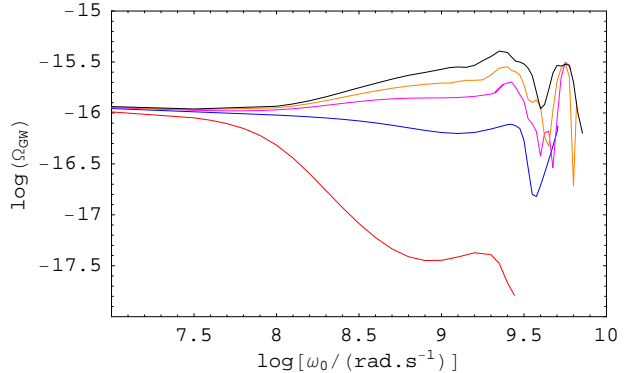}
\caption{Zoom of the high-frequency region of the spectra shown in
figure~\ref{fig-varios-n}.} \label{fig-varios-n-zoom}
\end{figure}

Gravitational-wave energy spectra, for different va\-lues of $p$,
are shown in Fig.~\ref{fig-varios-p}. For intermediate values of
the gravitational-wave frequency ($10^{-15}\mbox{ rad/s}\lesssim
\omega_0 \lesssim 10^{7}\mbox{ rad/s}$) the spectrum has a slope
that depends on the value of $p$. When $p$ approaches $-1$, the
model becomes closer to exponential inflation and the spectrum
closer to flat, as expected. For $\omega_0 \lesssim 10^{-15}\mbox{
rad/s}$, the energy spectrum is steeper than in the
intermediate-frequency region, as explained above. The behavior in
the high-frequency region of the spectra ($\omega_0 \gtrsim
10^7\mbox{ rad/s}$) can be understood as follows. Near its
minimum, the potential $V(\phi)$ is proportional to $\phi^n$. An
homogeneous scalar field, with a potential $V(\phi)\propto
\phi^n$, oscillating rapidly relatively to the expansion rate of
the universe, behaves as a fluid with an equation of state
$p=\gamma \rho$, $\gamma$ depending on $n$ in such a way that the
fluid becomes more and more stiff as $n$ increases \cite{turner}.
Furthermore, the relative logarithmic energy density of the
gravitational waves decreases at high-frequencies if inflation is
followed by a period dominated by a fluid with a dust-like
equation of state ($p=0$), while an increase of the relative
energy density is observed at high frequencies if inflation is
followed by a stiff matter phase ($p=\rho$) \cite{henriques}.
Since the spectra in Fig.~\ref{fig-varios-p} were obtained for
$n=2$, in which case the rapidly-oscillating scalar field behaves
like dust, a decrease at high frequencies of
$\Omega_{\mbox{\scriptsize GW}}$ is expected.

The dependence on the parameter $n$ of the relative logarithmic
energy density of the gravitational waves at high frequencies is
clearly illustrated in Figs.~\ref{fig-varios-n} and
\ref{fig-varios-n-zoom}, where the gravitational-wave energy
spectra for different values of $n$ are shown. The part of the
spectra corresponding to frequencies between $10^{7}\mbox{ rad/s}$
and $10^{10}\mbox{ rad/s}$ does reflect some of the
characteristics of the transition between the inflationary and
radiation-dominated eras, namely, an increase in the value of the
parameter $n$ corresponds to an increase of the relative
logarithmic energy density of the gravitational waves produced
during the period of thermalization near the minimum of the
potential.

Measurements of the cosmic microwave background radiation impose,
for angular frequencies corresponding to the present size of the
cosmological horizon $\omega_{\mbox{\scriptsize
hor}}=1.4\times10^{-17}\mbox{ rad/s}$, an upper limit on the
relative logarithmic energy density of gravitational waves,
namely, $\Omega_{\mbox{\scriptsize GW}} (\omega_{\mbox{\scriptsize
hor}},\eta_0) < 1.4 \times 10^{-10}$ \cite{allen}. All the spectra
presented in this paper satisfy, by far, this condition. In fact,
one could even relax the constraints on $p$ and $V_0$, discussed
in Sec.~\ref{sect-power-law}, and still obtain gravitational-wave
energy spectra satisfying the above constraint on
$\Omega_{\mbox{\scriptsize GW}} (\omega_{\mbox{\scriptsize
hor}},\eta_0)$.

\section{Conclusions}\label{sect-conclusions}

In this work, we have investigated the generation of gravitational
waves in power-law inflationary models. The parameters of the
inflationary potential were constrained using recent measurements
from the cosmic microwave background and large-scale structure. We
have incorporated reheating in our model through an elementary
decay mechanism of the inflationary scalar field into a
relativistic radiation fluid. Despite its simplicity, the
reheating potential (\ref{reheating-potential}) may contain
relevant features of more realistic potentials, as was emphasized
through an approximate comparison with the potential derived in
Ref.~\cite{anchordoqui-etal} from the Salam-Sezgin model.

We have used the method of continuous Bogoliubov coefficients to
calculated the gravitational-wave energy spectrum for different
values of the parameters of our model. All the spectra we have
obtained satisfy the constraint imposed by measurements of the
cosmic microwave background radiation at angular frequencies
corresponding to the present size of the cosmological horizon.
Between $10^{-15}\mbox{ rad/s}$ and $10^{7}\mbox{ rad/s}$ the
spectrum is not flat, but has a slope that depends on the value of
$p$, and so on the value of the parameter $\lambda$ defining the
inflationary potential (Fig.~\ref{fig-varios-p}). When $p$
approaches $-1$, the model becomes closer to exponential inflation
and the spectrum closer to flat, as expected. The influence of the
inflationary regime does extend a long way towards low frequencies
and the observation, or non-observation, of such a slope could, in
principle, give us information on the type of inflation. On the
other hand, the part of the spectrum corresponding to the
frequencies between $10^{7}\mbox{ rad/s}$ and $10^{10}\mbox{
rad/s}$ does reflect some of the characteristics of the transition
between the inflationary and radiation-dominated eras, in
particular the behavior during the period of thermalization near
the minimum of the potential, as explained at the end of the
preceding section, and shown in Figs.~\ref{fig-varios-n} and
\ref{fig-varios-n-zoom}. This is why we believe it is important to
continue work already done to develop detectors covering the MHz
and GHz range of frequencies: they might give us information
difficult to obtain by any other means. This was one point we
wanted to make with our present work, notwithstanding the fact
that the model we used was a very simple one.

\begin{acknowledgments}
This work was supported in part by the Funda\c{c}\~ao para a
Ci\^encia e a Tecnologia, Portugal.
\end{acknowledgments}

\end{document}